\documentclass[a4paper,10pt]{iopart}
\usepackage{iopams}
\expandafter\let\csname equation*\endcsname\relax
\expandafter\let\csname endequation*\endcsname\relax
\usepackage{amssymb,amsmath,graphicx}
\usepackage{multirow}  
\usepackage{booktabs} 
\usepackage{hyperref}
\usepackage{tikz}
\usetikzlibrary{arrows,shapes}

\numberwithin{equation}{section}

\newcommand{\Real}{\mathbb{R}}

\newcommand{\smfrac}[2]{\genfrac{}{}{}{1}{#1}{#2}}
\newcommand{\ZZ}{\mathbb{Z}}
\newcommand{\id}{\mathbf{1}}
\newcommand{\Hop}{H_{\text{op}}}
\newcommand{\Hcl}{H_{\text{per}}}
\newcommand{\hc}{\text{h.c.}}
\newcommand{\rrangle}{\rangle\!\rangle}
\newcommand{\llangle}{\langle\!\langle}
\newcommand{\frg}{g}
\newcommand{\Np}{n}
\newcommand{\ml}{m}
\newcommand{\hi}{h_{i,i+1}}
\newcommand{\hL}{h_{L,1}}

\begin{document}

\title[Topological and symmetry broken phases of $\ZZ_N$
parafermions in one dimension]{Topological and symmetry broken phases
  of\\$\ZZ_N$ parafermions in one dimension}


\author{Roberto Bondesan
 and Thomas Quella}
\address{
  Institute of Theoretical Physics, University of Cologne\\
  \small Z\"ulpicher Stra\ss{}e 77, D-50937 Cologne, Germany}
\eads{\mailto{Roberto.Bondesan@uni-koeln.de},
      \mailto{Thomas.Quella@uni-koeln.de}}

\date{}



\begin{abstract}
  We classify the gapped phases of $\ZZ_N$ parafermions in one
  dimension and construct a representative of each phase. Even in the
  absence of additional symmetries besides parafermionic parity,
  parafermions may be realized in a variety of phases, one for each
  divisor $\Np$ of $N$. The phases can be characterized by spontaneous
  symmetry breaking, topology, or a mixture of the two. Purely
  topological phases arise if $n$ is a unitary divisor, i.e.\ if $\Np$
  and $N/\Np$ are co-prime. Our analysis is based on the explicit
  realization of all symmetry broken gapped phases in the dual
  $\ZZ_N$-invariant quantum spin chains.
\end{abstract}

\section{Introduction}

  Topological phases of matter received a lot of interest recently due
  to their peculiar properties such as the existence of robust edge
  modes, non-abelian excitations and their potential applications to
  quantum computation
  \cite{Kitaev:1997QuantumComputation,Freedman:2000quant.ph..1108F}.
  While one-dimensional bosonic systems do not exhibit intrinsic
  topological order \cite{Chen:PhysRevB.83.035107}, the situation is
  very different for fermions. According to the general theory, there
  should exist two phases of fermions which are described by a
  $\ZZ_2$ topological invariant
  \cite{Ryu:1367-2630-12-6-065010,Kitaev:2009mg,Turner:2011PhRvB..83g5102T}.
  An explicit realization of both phases is provided by Kitaev's
  Majorana chain which models the dynamics of spinless fermions in a
  quantum wire, in proximity with a p-wave superconductor
  \cite{Kitaev:2001PhyU...44..131K}.
  
  The most remarkable feature of Kitaev's model is the emergence of
  stable gapless Majorana edge modes
  in the topologically non-trivial
  phase \cite{Kitaev:2001PhyU...44..131K}. In view of their non-abelian
  statistics, being able to create and to manipulate such isolated
  Majorana fermions would be a major step towards the realization of
  topological quantum computers \cite{Alicea:2011NatPh...7..412A}.
  While Kitaev's chain merely serves as
  a proof of principle, by now more realistic experimental setups for
  the creation and detection of Majorana fermions have been
  suggested \cite{Lutchyin:2010PhRvL.105g7001L,Oreg:2010PhRvL.105q7002O}.
  The basic ingredients are spin orbit quantum wires in proximity with
  an s-wave superconductor and in a magnetic field. Recent experiments
  based on this setup provide first evidence that Majorana fermions
  indeed exist \cite{Mourik:2012Sci...336.1003M,Das:2012Majorana} even
  though at least part of the observed features could also be
  explained by disorder
  effects
  \cite{Bagrets:2012PhRvL.109v7005B,Pientka:2012PhRvL.109v7006P,Neven:2013arXiv1302.0747N}.

  More recently, it has been investigated whether similar phenomena
  can be realized in heterostructures involving edges of {\em fractional} 
  topological insulators. It was found that such systems can be used
  to isolate parafermions which may be regarded as fractionalized
  Majorana fermions
  \cite{Clarke:2012arXiv1204.5479C,Lindner:2012arXiv1204.5733L,
    Cheng:2012arXiv1204.6084C,Vaezi:2012arXiv1204.6245V}.
  These efforts received their motivation from notes of P.~Fendley on
  the existence of edge modes in parafermionic chains,
  unpublished for a long time but eventually made available in
  Ref.~\cite{Fendley:2012arXiv1209.0472F}.
  Bulk lattice defects with parafermionic statistics
  emerged independently in the study of 
  fractional Chern insulators 
  \cite{Barkeshli:2012PhRvX...2c1013B}
  and $\ZZ_N$ rotor models 
  \cite{You:2012PhRvB..86p1107Y}, and 
  a general framework characterizing the statistics of all these defects
  was developed in 
  \cite{Barkeshli:2013PhRvB..87d5130B,Hastings:2012arXiv1210.5477H}.
  Furthermore,
  two-dimensional arrays of parafermions have been studied in
  \cite{Burrello:2013arXiv1302.4560B}.
  
  The study of parafermions has a long history. The exploration of
  para-particles began in
  Refs.~\cite{Green:PhysRev.90.270,Greenberg:PhysRev.138.B1155}
  in an attempt to generalize the canonical commutation relations of
  bosons and fermions. In contrast, these days the predominant notion
  of parafermions arises in the context of statistical physics, in
  connection with the so-called $\ZZ_N$ clock model
  \cite{Fradkin:1980th}. This model generalizes the $\ZZ_2$
  symmetry of the Ising model in a natural way to the $\ZZ_N$
  symmetry of a clock with $N$ marks.\footnote{The name
    ``parafermions'' is sometimes also used for the basic fields in a
    family of $\ZZ_N$ symmetric conformal field theories that
    arise at critical points of so-called clock models
    \cite{Fateev:1985mm}. These parafermions may be used to construct
    the Read-Rezayi states, i.e.\ they are of relevance for the study
    of fractional quantum Hall systems
    \cite{Read:1999PhRvB..59.8084R}. In the present paper, we shall be
    concerned exclusively with the gapped phases of $\ZZ_N$
    parafermions.}

  Just as the Ising model may be mapped to a problem of free fermions,
  using a variation of the non-local Jordan-Wigner transformation, the
  $\ZZ_N$ clock model exhibits a dual description in terms of
  fundamental parafermionic degrees of freedom. The parafermionic
  operators satisfy a $\ZZ_N$-graded variant of the usual
  Clifford algebra relations. This duality generalizes the well-known
  duality between the Ising model and free fermions for $N=2$. It is
  known that the topologically trivial and the non-trivial phase of
  the Majorana chain are equivalent to the ordered and the disordered
  phase of the Ising model, respectively
  \cite{Fidkowski:2011PhRvB..83g5103F}.
  
  The previous paragraphs suggest two natural avenues to the
  classification of parafermionic chains. One first of all can
  generalize the group theoretical approach of Ref.\
  \cite{Turner:2011PhRvB..83g5102T} that has been applied
  successfully to enumerate and characterize the different phases of
  Majorana chains with or without time-reversal symmetry. This
  approach has the advantage of yielding the maximal set of {\em
    possibilities} in a clean and conceptually clear way.
  On the other hand, the proof that all topological phases can
  actually be realized requires a method of combining phases such that
  their topological charges add up. This procedure might not
  necessarily lead to the most natural representative of the
  new phase.
  Moreover, the approach does not reveal the possible patterns
  of symmetry breaking. This approach has been followed 
  recently in Ref.~\cite{Motruk:2013arXiv1303.2194M}
  for purely topological phases.

  The second approach which shall be followed in this paper is a
  complete characterization of gapped symmetric and symmetry broken
  phases of $\ZZ_N$ quantum spin chains. These can be engineered
  systematically on the level of ground states and their
  associated frustration free parent Hamiltonians
  \cite{Schuch:1010.3732v3}. Via the duality mapping mentioned before,
  this classification provides an explicit characterization of all
  parafermionic phases which cannot be connected without crossing a
  phase transition, i.e.\ closing the gap.

  We shall find that the distinct parafermionic phases are labeled by
  the possible divisors of $N$. Generically, they may exhibit a
  coexistence of both ferromagnetic and topological order. In
  practice, this means that only part of the ground state degeneracy
  of an open chain is topologically protected and due to gapless edge
  modes, while the remaining degeneracy can be lifted by spontaneous
  symmetry breaking. When the chain is closed, only the topological 
  degeneracy disappears, but a non-trivial ground state degeneracy
  persists in general. To be more specific, for a given divisor $\Np$
  of $N$, the phase will be purely topological if $N$ and $N/\Np$ are
  co-prime, and it will exhibit a unique ground state in a closed
  system and gapless edge modes in an open system. If instead
  $N/\Np=\gcd(\Np,N/\Np)$ the phase is topologically trivial and
  entirely characterized by the existence or absence of symmetry
  breaking. The details of the classification will be explained in the
  main text through the concrete realization of each individual
  phase.

  We would like to stress that our classification distinguishes only
  phases which cannot be connected by any local $\ZZ_N$-invariant
  interaction. Clearly, specific models with an explicit Hamiltonian
  depending on a given set of coupling constants may have arbitrarily
  complicated phase diagrams.
  
  The paper is organized as follows. In Section~\ref{sc:Setup} we
  introduce the kinematical setup that we are working with. This
  includes a detailed discussion of the symmetry algebra and of the
  Hilbert spaces in question. The possible symmetry breaking patterns
  and order parameters of the $\ZZ_N$ quantum spin chain are analyzed in
  Section~\ref{sc:SSB} and a representative Hamiltonian is constructed
  for each phase. These results form the heart of our paper. Finally,
  we perform a non-local duality transformation in
  Section~\ref{sc:PhasesPF} and interpret our
  previous results from a parafermionic perspective. As it turns out,
  the original symmetry broken phases of the spin chains generally
  exhibit a mixture of symmetry breaking and topological
  protection. We end with concluding remarks and an outlook to open
  problems.
 
\section{\label{sc:Setup}Parafermions and $\ZZ_N$ 
spin chains}

  We introduce the mathematical framework within which our
  considerations take place. This includes a detailed discussion of
  the symmetry algebra and of the Hilbert spaces involved.

\subsection{$\ZZ_N$ spin chains}

  $\ZZ_N$ quantum spin chains are $\ZZ_N$ symmetric models
  where the spins can show in any of $N$ directions in the plane which
  can be thought of as corresponding to the $N$ distinct $N^{\text{th}}$
  roots of unity on the unit circle. This setup generalizes the
  familiar $\ZZ_2$ case of the Ising model where one deals with 
  spins pointing either up or down, and at the same time, provides a
  specialization of the $XY$ model in which the spins are restricted
  to take discrete values.

  If we denote the size of the chain by $L$, the Hilbert space of the
  chain will be given by
\begin{equation}
  \mathcal{H}=(\mathbb{C}^N)^{\otimes L}\, .
\end{equation}
We choose a basis in $\mathcal{H}$ given by $|q_1,\dots,q_L\rangle$,
$q_i\in \ZZ_N$. (This means that $|q\rangle\equiv |q+N\rangle$.)
Define the operators 
\begin{align}
  \sigma_i&=\id\otimes \cdots \otimes \underbrace{\sigma}_{i}\otimes \id \cdots
  \otimes\id\\ 
\tau_i&=\id\otimes \cdots \otimes  \underbrace{\tau}_{i}\otimes \id \cdots \otimes\id\, ,
\end{align}
where $\sigma$ and $\tau$ are
\begin{equation}
  \sigma =
  \begin{pmatrix}
    1&0&0&\dots&0\\
    0&\omega&0&\dots&0\\
    0&0&\omega^2&\dots&0\\
    \vdots&\vdots&\vdots&\ddots&\vdots\\
    0&0&0&0&\omega^{N-1}
  \end{pmatrix}
\, ,\quad
  \tau =
  \begin{pmatrix}
    0&1&0&\dots&0\\
    0&0&1&\dots&0\\
    \vdots&\vdots&\ddots&\vdots&\vdots\\
    0&\dots&0&0&1\\
    1&\dots&0&0&0
  \end{pmatrix}\, .
\end{equation}
  $\sigma,\tau$ satisfy the relations:\footnote{
  These are the relations of the Heisenberg group which is defined by
  exponentiation of the Weyl algebra $[Q,P]=QP-PQ=i\id$. In terms of
  the generators $\sigma=\exp(i P)$, $\tau=\exp(i Q)$ and the central
  element $Z=\exp(i\id)$ the group multiplication reduces to
  $\sigma^r\tau^s=Z^{rs}\tau^s\sigma^r$ for any parameters $r,s\in\Real$.
}
\begin{align}
  X^{N}&=1\, , X^{\dagger}=X^{-1}\, ,\quad
  \text{for } X=\sigma,\tau\\
  \sigma^r\tau^s&=\omega^{rs}  \tau^s\sigma^r\, ,
   \quad \omega = e^{2\pi i/N}\, ,
 \quad  r,s\in \ZZ_N\, ,
\end{align}
and as obvious from the definition, $\sigma_i,\tau_j$ commute if $i\neq j$.
Note that for $N=2$: $\sigma=\sigma^z$, $\tau=\sigma^x$.
The space of linear operators on $\mathcal{H}$ is spanned
by monomials $\prod_{i=1}^L\sigma_i^{p_i}\tau_i^{q_i}$,
$p_i,q_i\in \ZZ_N$.

The global symmetry $G=\ZZ_N$ is generated by
the operator shifting the labels by one unit on every site:
\begin{align}
\label{eq:Ptau}
  P = \prod_{i=1}^L\tau_i^\dagger\, ,
\end{align}
and each tensorand $\mathbb{C}^N$ in 
the Hilbert space $\mathcal{H}$ is direct sum of $N$ one-dimensional
irreducible representations of $\ZZ_N$.
The Hamiltonians $H$ we are going to consider will be invariant under
the action of $\ZZ_N$. In other words, $[H,P]=0$.

\subsection{Jordan Wigner transformation}

Now we perform a generalized Jordan Wigner transformation
to define the parafermions \cite{Fradkin:1980th}.
First we define operators living on the dual lattice as:
\begin{align}
\label{eq:dual}
  \mu_{i+1/2}=\sigma_i \sigma_{i+1}^\dagger\, ,\quad
  \rho_{i+1/2}=\prod_{k=1}^i\tau_k^\dagger \, ,
\end{align}
so that $\tau_i = \rho_{i-1/2}\rho_{i+1/2}^\dagger$, and $\mu,\rho$
satisfy the same commutation relations as respectively $\tau,\sigma$.
$\rho$ is the disorder operator dual to the order operator $\sigma$,
implementing the action of symmetries on spins between sites $1$ and
$i$, namely shifting those spins by one unit. Parafermions are then
defined on middle points between
vertices of the original and dual lattice,
as the product of order and disorder variables:
\begin{align}
  \label{eq:TrafoPFSpin}
  \gamma_{2i-1}=\left(\prod_{k=1}^{i-1}\tau_k \right)\sigma_i
  \, ,\quad
  \gamma_{2i} 
  =\omega^{(N-1)/2}
  \left(\prod_{k=1}^{i-1}\tau_k \right)\sigma_i\tau_i\, .
\end{align}
They satisfy the following commutation relations
\begin{align}
  \gamma_i^N=1\, ,\quad
  \gamma_i\gamma_j=\omega \gamma_j\gamma_i\, \quad(\text{for } 1\le i<j\le 2L)\, ,
\end{align}
  and the definition can be inverted to yield:
\begin{align}
  \label{eq:TrafoSpinPF}
  \sigma_i\sigma_{i+1}^\dagger
  &=\omega^{(N+1)/2} \gamma_{2i}\gamma^{-1}_{2i+1}\\
  \tau_i
  &= \omega^{-(N-1)/2} \gamma_{2i-1}^{-1}\gamma_{2i}\, .
\end{align}

$\gamma$'s are the generators of an algebra which generalize the
Clifford algebra of ``Majorana fermions'' occurring at $N=2$
\cite{Morris1967}. Elements of a generalized Clifford algebra (with even number
of generators, as in the case at hand, where the algebra is isomorphic
to the complex matrices acting on the Hilbert space of the chain) are
the sum of homogeneous elements $X$ which have a well defined
$\ZZ_N$-grading $|X|\in \ZZ_N$ given by:
\begin{align}
  P X P^{-1} = \omega^{|X|} X\, ,
\end{align}
with $P$ the symmetry generator expressed in terms of the parafermions as
\begin{equation}
  P=\prod_{i=1}^L \omega^{(N-1)/2} \gamma_{2i}\gamma_{2i-1}^{-1} \, .
\end{equation}
  In particular, the parafermionic generators themselves are elements
  of degree $1$.

\section{Symmetry broken phases in $\ZZ_N$ spin chains}
\label{sc:SSB}\label{sec:ssb}

The general theory developed in
\cite{Schuch:1010.3732v3,Chen:PhysRevB.84.235128} classifies
gapped phases of spin chains with symmetry group $G$ in terms of
classes of projective representations of subgroups $H$ to which the
symmetry is spontaneously broken.  The subgroups of $\ZZ_N$ are given
by $\ZZ_{\Np}$, where $\Np$ is a divisor of $N$. $\ZZ_N$ only has
trivial projective representations for any $N$,
since the phase $e^{i\phi}$ appearing in the projective version of the 
defining relation, $P^N=e^{i\phi} \id$,
can always be absorbed in the definition of $P$.
Therefore gapped phases of $\ZZ_N$ spin
chains are indexed by divisors of $N$ and fully characterized by
spontaneous symmetry breaking (SSB).

Statistical systems with $\ZZ_N$ symmetry have been studied
extensively over the years, and their phase diagrams
have been discussed in a variety of places, see
e.g.~\cite{Ostlund1981,Alcaraz1981,Fateev:1985mm,Ortiz2012}.  In this
section, we explicitly construct representatives of all gapped phases we
expect to have. On the way, we shall emphasize important aspects
related to order parameters which will play a key role in the
description of parafermionic phases in Section \ref{sc:PhasesPF}.

\subsection{Construction of the phases}

We first briefly recall the defining features of SSB in $\ZZ_N$ spin
chains. The breaking of $\ZZ_2$ symmetry in the ordered phase of
the quantum Ising chain is an elementary textbook example
\cite{Sachdev2001}: the phase is characterized by a non-vanishing
expectation value of the order parameter $\sigma$.
The generalization to the breaking of $\ZZ_N$
down to $\ZZ_{\Np}$, $\Np$ divisor of $N$, is straightforward.

In the thermodynamic limit, the local order parameters describing the
properties of the spin chain are given by the powers $\sigma^{\alpha
  \Np}$ with $\alpha\in\ZZ_\ml\backslash\{0\}$ and $\ml=N/\Np$. The
group of residual
symmetries which leave the order parameter invariant is 
$\{1,P^m,\dots,P^{(n-1)m}\} = \ZZ_{\Np}$. It
is generated by $P^{\ml}$, the $\ml^{\text{th}}$ power of the global
parity.
(If $\Np=N$ the phase is disordered.)
Non-trivial elements of 
$\ZZ_N/\ZZ_{\Np}=\{1,P,\dots,P^{m-1}\}=\ZZ_\ml$ 
permute the ground states cyclically among themselves, so that
the ground state manifold is $\ml$--fold degenerate.  
We will refer to $\ZZ_\ml$ as the broken symmetries. The
degeneracy of the ground state encodes a permutation representation
of the symmetries on the ground state manifold. This representation
corresponds to an invariant of the phase and cannot change unless a
phase transition occurs \cite{Schuch:1010.3732v3}.  The symmetry breaking
perturbation $h\sum_i(\sigma_i^{\beta \Np}+\hc)$ ($h$ small and
positive) of the Hamiltonian will select a single ordered ground state,
leading to a non-zero expectation value of the order parameter 
 even when the external field $h$ is set to zero.

For later convenience we will now explicitly construct representatives
of each phase. This includes both the desired states and an associated
parent Hamiltonian. According to the above discussion, a phase labeled
by $\Np$ should have $\ml$ ordered ground states $|\psi_\alpha\rangle$
($\alpha\in\ZZ_\ml$) which satisfy
\begin{equation}
  P|\psi_{\alpha}\rangle
  =|\psi_{\alpha+1}\rangle \, ,\quad
  P^{k \ml}|\psi_{\alpha}\rangle
  =|\psi_{\alpha}\rangle \, ,
\end{equation}
for all $k\in\ZZ_{\Np}$.
  A simple solution of these constraints is given by the product
  states
\begin{equation}
  \label{eq:psi_def}
  |\psi_\alpha\rangle 
  =\bigotimes_{i=1}^L|\alpha\rrangle_i\, ,
\quad\text{ where }\quad 
  |\alpha\rrangle :=\frac{1}{\sqrt{\Np}} 
  \sum_{k=0}^{\Np-1} |\alpha+\ml k\rangle\, .
\end{equation}
  The states $|\alpha\rrangle$ are orbits under the subgroup
  $\ZZ_{\Np}$ of $\ZZ_N$. As such they span the subspace
  $\mathbb{C}^{\ml}\subseteq \mathbb{C}^N$ and they satisfy the
  periodicity condition $|\alpha\rrangle\equiv |\alpha+\ml\rrangle$.

  The associated Hamiltonian will be defined in such a way that the
  $\ml$-fold degeneracy of the ground state is already exact in finite
  systems and not only in the thermodynamic limit. This is achieved by
  choosing a frustration free combination of specific projectors which
  are localized on two neighboring sites. Taking $H=-\sum_i \hi$, the
  following choice of the two-body term $\hi$ guarantees the
  $\ZZ_N$-invariance and the projector property ($\hi^2=\hi$,
  $\hi^\dagger=\hi$)
\begin{equation}
  \label{eq:HamiltonianSpin}
  \hi=
  \sum_{\alpha=0}^{\ml-1}
  |\alpha\rrangle_i|\alpha\rrangle_{i+1}
  \llangle \alpha|_{i} \llangle\alpha|_{i+1}\, .
\end{equation}
The Hamiltonian $H$ is a sum of commuting projectors, so that it is
possible to diagonalize all terms simultaneously. The resulting spectrum is
given by the integers between $-L$ and $0$, and hence the chain is
gapped as desired. Assuming periodic or open boundary
conditions will make no difference for the conclusions we want to draw
in this section. In both cases the space of ground states is spanned
by the vectors $|\psi_\alpha\rangle$ with $\alpha\in
\ZZ_{\ml}$, for which $\hi=1$.

  It remains to find a more explicit form of the Hamiltonian
  \eqref{eq:HamiltonianSpin} in terms of the spin operators
  $\sigma_i$ and $\tau_i$. Using the single-site matrix elements
\begin{align}
  \langle q |\alpha\rrangle \llangle \alpha| p\rangle
  = 
  \frac{1}{N} \langle q |
  \sum_{\beta=0}^{\ml-1}(\sigma \omega^{-\alpha})^{\Np \beta} 
  \sum_{j=0}^{\Np-1}\tau^{\ml j}
  | p\rangle\, ,
\end{align}
  it can easily be verified that Eq.~\eqref{eq:HamiltonianSpin}
  reduces to
\begin{align}
  \label{eq:hiip1}
  \hi 
  =\frac{1}{\ml \Np^2}
  \sum_{\beta=0}^{\ml-1}(\sigma_i\sigma^\dagger_{i+1})^{\beta \Np} 
  \sum_{j,k=0}^{\Np-1}\tau_i^{j\ml}\tau_{i+1}^{k\ml}\, .
\end{align}
  We have thus succeeded in constructing gapped Hamiltonians realizing
  each possible SSB phase of $\ZZ_N$ spin chains.

  Before moving on, we would like to comment on the specific form of
  the Hamiltonians obtained. The cases $\Np=1$ and $\Np=N$ correspond,
  respectively, to the situations with no residual symmetry and full
  symmetry of the ground state. In these two extreme cases, the
  interaction simplifies to the expressions
\begin{align}
  (\Np,\ml)=(1,N)\, : \, \quad \hi &= \frac{1}{N}
  \sum_{\alpha=0}^{N-1}(\sigma_i\sigma^\dagger_{i+1})^{\alpha}\\
  (\Np,\ml)=(N,1)\, : \, \quad \hi &= \frac{1}{N^2}
  \sum_{j,k=0}^{N-1}\tau_i^{j}\tau_{i+1}^{k}\, .
\end{align}
  It is now evident that the general formula \eqref{eq:hiip1}
  is obtained by multiplying together the Hamiltonians of a
  $\ZZ_{\ml}$ chain whose symmetry is completely broken, $1/\ml
  \sum_{\beta=0}^{\ml-1}(\sigma_i^{\Np}
  (\sigma^\dagger_{i+1})^{\Np})^{\beta}$, and that of a $\ZZ_{\Np}$
  chain without symmetry breaking, $1/\Np^2
  \sum_{j,k=0}^{\Np-1}(\tau_i^\ml)^{j}(\tau_{i+1}^\ml)^{k}$.

  While our definition reproduces the conventional form of the
  ferromagnetic clock model deep in the symmetry broken phase
  \cite{Fendley:2012arXiv1209.0472F} for $\Np=1$, the case $\Np=N$
  leads to a more complicated expression than the one used in
  \cite{Fendley:2012arXiv1209.0472F}, namely
  $\sum_{j=0}^{N-1}\tau_i^{j}$
(even though both Hamiltonians have the same symmetric ground
state). Moreover, the duality transformations \eqref{eq:dual} between
the ordered and the disordered phase do not preserve the structure of
the Hamiltonians~\eqref{eq:hiip1}. It is therefore reasonable to
search for a slight modification of the Hamiltonian~\eqref{eq:hiip1}
such that it reduces to the form of the spin chains considered in
\cite{Fendley:2012arXiv1209.0472F}, while retaining the ground states
above. This goal is achieved by considering the expression
\begin{equation}
  \label{eq:SpinHamiltonianNew}
  \hat{H}
  =
-  \sum_i\left(
    \frac{1}{\ml}
    \sum_{\beta=0}^{\ml-1}(\sigma_i\sigma^\dagger_{i+1})^{\beta \Np} +
  \frac{1}{\Np}\sum_{j=0}^{\Np-1}\tau_i^{j\ml}\right)\, .
\end{equation}
This operator is Hermitian, $\ZZ_N$ invariant, 
and each single term is a projector commuting with all the other
terms. Therefore the chain is still gapped after the modification,
albeit with a different spectrum in comparison to~$H$.

The ground states are such that each term
is $1$ and $\hat{H}$ has exactly $\ml$ ground states which are
given by \eqref{eq:psi_def}.
Indeed, the terms $1/\Np \sum_{j=0}^{\Np-1}\tau_i^{j\ml}$ act as identity
on states $\bigotimes_{i=1}^L|\alpha_i\rrangle_i$
($\alpha_i\in\ZZ_\ml$). Demanding, in addition, that also
$1/\ml \sum_{\beta=0}^{\ml-1}(\sigma_i\sigma^\dagger_{i+1})^{\beta \Np}$
acts as identity selects states with $\alpha_i=\alpha_{i+1}$.
Hence $\hat{H}$ is another representative of the SSB phase under consideration.
Note that the Hamiltonians $\hat{H}$ in \eqref{eq:SpinHamiltonianNew}
admit a simple action (in the thermodynamics limit) under the duality
transformations~\eqref{eq:dual}. The order-disorder duality simply
swaps the phases labeled by $\Np$ with those labeled by $\ml$, and at
the same time interchanges $\sigma$ with $\rho$ and $\tau$ with
$\mu$. Nonetheless, we will stick to our original definition of the
Hamiltonian, see Eq.~\eqref{eq:hiip1}, in what follows.

\subsection{\label{sc:OP}Order parameters}

In a macroscopic system with broken symmetry one will not observe a
quantum superposition of ground states. Fluctuations will rather
single out a unique ground state with a well-defined magnetization. In
order to select a single ground state $|\psi_0\rangle$, we add a
symmetry breaking perturbation $h\sum_i(\sigma_i^{\beta\Np}+\hc)$ to
the Hamiltonian \eqref{eq:hiip1}. Afterwards we take the thermodynamic
limit and set $h=0$. It is easy to check that this procedure implies a
spontaneous magnetization
\begin{equation}
  \label{eq:av_sigma}
  \bigl\langle \sigma_i^{\beta \Np}\bigr\rangle := \langle \psi_0|
  \sigma_i^{\beta \Np}|\psi_0\rangle\neq 0 \, ,
\end{equation}
and also 
$\langle\tau_{i}^{k\ml}\rangle \neq 0$ for all $k\in\ZZ_{\Np}$. More
importantly, we also have
\begin{equation}
  \label{eq:av_rho}
  \bigl\langle\rho_{i+1/2}^{k\ml}\bigr\rangle \neq 0\, .
\end{equation}
In contrast, all other powers of 
$\sigma$ and $\rho$ have vanishing expectation values.
Now we expect that
slight modifications of Hamiltonians of gapped phases cannot change
abruptly the expectation values unless a phase transition is reached.
Then Eqs.~(\ref{eq:av_sigma},~\ref{eq:av_rho}) hold
not only for the fine tuned Hamiltonians we have constructed but
in a whole region of non-zero measure in the phase diagram, which
constitutes the SSB phase. 

The mechanism of symmetry breaking is well understood
for the Ising chain and it
applies similarly to $\ZZ_N$ spins. The operator $\rho_{i+1/2}$
disorders an ordered state by creating kink excitations, topological
objects rotating all spins from site $1$ to $i$ by one unit. The
disordered phase can be characterized by the condensation of these
kinks \cite{Fradkin1978}.

We are now prepared to appreciate that a particular situation is encountered when there
exist two integers $\alpha,k$ such that $\alpha \Np = k \ml$. Indeed,
under these circumstances we are able to construct another
{\em non-local} order parameter 
\begin{equation}
  \label{eq:OPNew}
  (\rho_{i-1/2}^\dagger \sigma_i)^{\alpha \Np}\, .
\end{equation}
(The non-zero expectation value in the state $|\psi_0\rangle$
is implied by Eqs.~(\ref{eq:av_sigma},~\ref{eq:av_rho}).) With the definition $\frg:=\gcd(\Np,\ml)$, the solutions of the
equation $\alpha\Np=k\ml$ can be parametrized as follows,
\begin{align}
  \alpha = k \frac{\ml}{\Np}\,\quad\text{ with }\quad 
  k\in \Bigl\{0, \smfrac{\Np}{\frg}, 2\smfrac{\Np}{\frg}, 
  \dots, (\frg-1)\smfrac{\Np}{\frg}\Bigr\}= \smfrac{\Np}{\frg} \ZZ_{\frg}\, .
\end{align}
  We conclude that we have additional (non-local) order parameters
  whenever $\frg>1$. Their existence is a direct consequence of being
  in a SSB phase of the $\ZZ_N$ spin chains. (Of course the presence
  of non-local order parameters does not mean that the
  phase at hand is topological, since the degeneracy of the ground
  state of the spin chain
  can be completely lifted through SSB.) In Section
  \ref{sec:par_phases}, we will shed more light onto the physical
  implications of the presence of this new non-local order
  parameter. It is no coincidence that the product \eqref{eq:OPNew} of
  order and disorder variables that we have just encountered agrees
  with powers of the parafermion, see Eq.~\eqref{eq:TrafoPFSpin}.

\section{\label{sc:PhasesPF}Classification of parafermionic phases}

  In this section we will translate the classification of symmetry
  breaking patterns in $\ZZ_N$ spin chains into the dual
  language of parafermions. It will be found that the resulting
  parafermionic Hamiltonians generally exhibit a mixture of symmetry
  breaking and non-trivial topology which can be made manifest in the
  ground state degeneracies, in the occurrence of protected edge
  modes and in the possibility to define an integer valued topological
  invariant. We shall point out essential differences in the ground
  state degeneracy that arise when considering closed boundary
  conditions as opposed to open ones.

\subsection{Duality transformation of the Hamiltonian}

  Our first goal is to rewrite the Hamiltonian Eq.~\eqref{eq:hiip1} in
  terms of the parafermions. Using the transformation rules
  \eqref{eq:TrafoSpinPF} and the relation
\begin{align}
  (\gamma_i^{a} \gamma^b_j)^{c}=\omega^{-abc(c-1)/2}
  \gamma_i^{ac} \gamma_j^{bc}\quad(\text{for } i<j)\, ,
\end{align}
  we immediately obtain the two-site Hamiltonian
\begin{align}
  \label{eq:hiip1_para}
  \begin{split}
  \hi&=\frac{1}{\ml}\sum_{\beta=0}^{\ml-1}
    \omega^{\beta \Np(N + \beta \Np)/2} 
  \gamma_{2i}^{\beta \Np}\gamma^{-\beta \Np}_{2i+1}\\ 
&\qquad\qquad \times\frac{1}{\Np^2}  
  \sum_{j,k=0}^{\Np-1}
  \omega^{(\ml^2(j^2+k^2)-N\ml (j+k))/2}
  \gamma_{2i-1}^{-j\ml}
  \gamma_{2i}^{j\ml} 
  \gamma_{2i+1}^{-k\ml}\gamma_{2i+2}^{k\ml}\, .
  \end{split}
\end{align}
Despite the non-locality of the transformation, the Hamiltonian still
remains local when expressed in terms of the parafermionic
variables. The only complication arises for periodic boundary
conditions since the naive application of the Jordan Wigner
transformation generates a non-local term in this case. In other
words, locality implies that the Hamiltonian for the periodic
parafermionic chain needs to be different from that of the periodic
spin model, see Section \ref{sc:ClosedBC} for details.

In the following we will use a basis of parity eigenvectors 
in the space of ground states which is more natural to work with 
in the parafermionic picture:
\begin{align}
  \label{eq:psia}
  |\phi_\alpha\rangle := \sum_{\beta=0}^{\ml-1}\omega^{\beta \alpha \Np}
    |\psi_\beta\rangle\, ,
\end{align}
so that 
\begin{equation}
  \label{eq:parity_gs}
  \sigma_i^{\beta \Np} |\phi_\alpha\rangle =
  |\phi_{\alpha+\beta}\rangle\, ,\quad
  \tau_i^{k \ml} |\phi_\alpha\rangle =
  |\phi_{\alpha}\rangle\, ,\quad
  P  |\phi_\alpha\rangle = \omega^{-\alpha \Np}|\phi_\alpha\rangle\, .  
\end{equation}
  
\subsection{\label{sec:par_phases}Quantum phases}

  In Section \ref{sec:ssb} we characterized SSB phases of $\ZZ_N$
  spin chains. However, the local order parameter $\sigma^{\alpha
    \Np}$ of the spin chain becomes non-local when expressed in terms
  of parafermionic variables. Since physical perturbations are assumed
  to be local, this observation suggests that there is no way to
  spontaneously break the symmetry in the parafermionic chain, and
  that the phases are characterized by topological order instead of
  ferromagnetic order. In this case, the states emerging at the
  macroscopic level would be those with definite $\ZZ_N$
  parity. In the fermionic case $N=2$, for example, it is
  well-established that there is no local order parameter, and
  correspondingly no local perturbation which can break the symmetry
  \cite{Turner:2011PhRvB..83g5102T,Fidkowski:PhysRevB.81.134509}.

  While valid for a considerable subset of symmetry breaking patterns
  in the $\ZZ_N$ spin chains, the previous expectation fails to
  be true in general. This exactly happens in the phases labeled by a
  divisor $\Np$ such that $\frg=\gcd(\Np,N/\Np)>1$. Also in the dual
  picture, parafermionic parity can be broken (partly) in these cases
  and, depending on the precise values of $\frg$ and $\Np$, this leads
  to phases which, in addition to ferromagnetic order, might as well
  exhibit topological order.

In order to substantiate our claim, let us briefly
recall our results from Section~\ref{sc:OP}. For $\frg>1$ we have
pointed out the existence of a second set of order parameters
\eqref{eq:OPNew}, which  become {\em local} when rewritten in terms of
parafermions. More precisely, they correspond exactly to mutually
commuting powers of the parafermions,
\begin{align}
  \Bigl[\gamma_{i}^{p N/\frg},
    \gamma_{j}^{p'N/\frg}\Bigr]=0
  \quad(\text{with }p,p'\in \ZZ_\frg)\, .
\end{align}
Note that, more generally,  $\gamma_{j}^{p N/\frg}$
commutes with $\gamma_{i}^{\alpha \Np}$ and $\gamma_{i}^{k\ml}$
for every $i,j$, so that
\begin{align}
\label{eq:comm_gamma_nm}
  \Bigl[\gamma_{j}^{p N/\frg},\hi \Bigr]=0 \, ,
\end{align}
  implying a non-trivial ground state degeneracy.
  The parafermionic local order parameters transform non-trivially
  under the action of elements in the set $\{P,\dots,P^{g-1}\}\subset
  \ZZ_N$. Hence they can be used to detect a symmetry breaking from
  $\ZZ_N$ to $\ZZ_{N/\frg}$.

  To show explicitly how the symmetry breaking occurs we rely on the
  Hamiltonian \eqref{eq:hiip1_para}. We suppose open boundary
  conditions, for which the Hamiltonians for the spins and the
  parafermions are identical. Let us then consider a $\ZZ_N\to
  \ZZ_{N/\frg}$ symmetry breaking perturbation of the original
  parafermionic Hamiltonian,
\begin{equation}
  \label{eq:SSB}
  H\,\mapsto\,
  H+ h \sum_i\Bigl\{\gamma_{i}^{p N/\frg}+\hc\Bigr\}\quad
  (\text{with }p\in \ZZ_\frg\backslash\{0\})\, .
\end{equation}
  The extra term in the Hamiltonian is local in the parafermionic
  variables. It acts on the ground states as
\begin{align}
 \gamma_{2i-1}^{p N/\frg} |\psi_\alpha  \rangle
  = 
  \left(\,\prod_{k=1}^{i-1}\tau_k \right)^{p N/\frg} 
  \sigma_i^{p N/\frg} |\psi_\alpha  \rangle
  = \omega^{\alpha p N/\frg}
  |\psi_{\alpha}  \rangle
  \, ,
\end{align}
  and analogously for $\gamma_{2i}$, with an extra
  $\alpha$-independent phase. As a consequence, the SSB induced by the
  perturbation \eqref{eq:SSB} singles out a subset of the $\ml$ ground
  states, namely exactly those labeled by an element $\alpha$ such
  that
\begin{equation}
  \alpha\,\frac{p N}{\frg}=0\pmod{N}\, .
\end{equation}
  It is evident that this equation has $\ml/\frg$ solutions
\begin{equation}
  \alpha \in \{0,\frg,\dots,\ml-g\} =\frg \ZZ_{\ml/\frg}\,  .
\end{equation}
  Taking $h\to 0$ after passing to the thermodynamic limit gives
  $\langle \psi|\gamma_i^{p N/\frg}|\psi\rangle=
\langle \psi|\psi\rangle\neq 0$, where $|\psi\rangle$ belongs to the 
$\ml/\frg$-dimensional space
of residual ground states spanned by $\{|\psi_0\rangle,
|\psi_\frg\rangle,\dots,|\psi_{\ml-\frg}\rangle\}$.
In Section \ref{sc:Realization} we will show that the residual
degeneracy is associated with the presence of edge modes and should
  therefore be regarded as topological.
If however $\ml=\frg$ no such degeneracy is present, as e.g.~when
$N=4$ and $\Np=2$, and the system is in a purely symmetry breaking
phase.

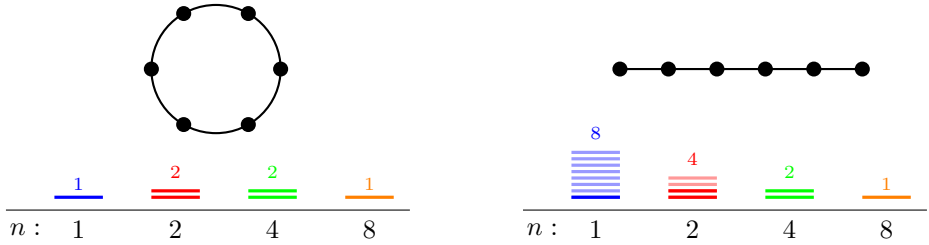
\begin{figure}[htp]
  \centering
\begin{tikzpicture}[scale=0.85,thick]
  \begin{scope}[xshift=6.25cm]
  \draw[] (0,0)--(3.75,0);
  \foreach \x in {0,0.75,...,3.75} 
  { 
    \draw[fill=black] (\x,0) circle (0.1cm);
  }
  \begin{scope}[yshift=-2cm,xshift=-1cm]
    \draw[thin] (-0.5,-0.2)--(5.75,-0.2);
    \node[] at (-0.2,-0.5) {$\Np:$};
  \begin{scope}[xshift=0.25cm]
    \node[] at (0.375,-0.5) {$1$};
    \node[blue] at (0.375,1) {{\scriptsize $8$}};
    \foreach \y in {0,0.1,...,0.8} 
    { 
      \draw[blue!40!white,very thick] (0,\y)--(0.75,\y);
    }
      \draw[blue,very thick] (0,0)--(0.75,0);
  \end{scope}
  \begin{scope}[xshift=1.75cm]
    \node[] at (0.375,-0.5) {$2$};
    \node[red] at (0.375,0.6) {{\scriptsize $4$}};
    \foreach \y in {0,0.1,...,0.4} 
    { 
      \draw[red!40!white,very thick] (0,\y)--(0.75,\y);
    }
      \draw[red,very thick] (0,0)--(0.75,0);
      \draw[red,very thick] (0,.1)--(0.75,.1);
  \end{scope}
  \begin{scope}[xshift=3.25cm]
    \node[] at (0.375,-0.5) {$4$};
    \node[green] at (0.375,0.4) {{\scriptsize $2$}};
    \foreach \y in {0,0.1,...,0.2} 
    { 
      \draw[green,very thick] (0,\y)--(0.75,\y);
    }
  \end{scope}
  \begin{scope}[xshift=4.75cm]
    \node[] at (0.375,-0.5) {$8$};
    \node[orange] at (0.375,0.2) {{\scriptsize $1$}};
    \foreach \y in {0} 
    { 
      \draw[orange,very thick] (0,\y)--(0.75,\y);
    }
  \end{scope}
  \end{scope}
  \end{scope}
  \begin{scope}[xshift=0cm]
    \draw[] (0,0) circle (1cm);
    \foreach \x in {0,1,2,3,4,5} 
    { 
      \draw[fill=black] ({cos(\x*360/6)},{sin(\x*360/6)}) circle (0.1cm);
    }
  \begin{scope}[yshift=-2cm,xshift=-2.75cm]
    \draw[thin] (-0.5,-0.2)--(5.75,-0.2);
    \node[] at (-0.2,-0.5) {$\Np:$};
  \begin{scope}[xshift=0.25cm]
    \node[] at (0.375,-0.5) {$1$};
    \node[blue] at (0.375,0.2) {{\scriptsize $1$}};
    \foreach \y in {0} 
    { 
      \draw[blue,very thick] (0,\y)--(0.75,\y);
    }
  \end{scope}
  \begin{scope}[xshift=1.75cm]
    \node[] at (0.375,-0.5) {$2$};
    \node[red] at (0.375,0.4) {{\scriptsize $2$}};
    \foreach \y in {0,0.1,...,0.2} 
    { 
      \draw[red,very thick] (0,\y)--(0.75,\y);
    }
  \end{scope}
  \begin{scope}[xshift=3.25cm]
    \node[] at (0.375,-0.5) {$4$};
    \node[green] at (0.375,0.4) {{\scriptsize $2$}};
    \foreach \y in {0,0.1,...,0.2} 
    { 
      \draw[green,very thick] (0,\y)--(0.75,\y);
    }
  \end{scope}
  \begin{scope}[xshift=4.75cm]
    \node[] at (0.375,-0.5) {$8$};
    \node[orange] at (0.375,0.2) {{\scriptsize $1$}};
    \foreach \y in {0} 
    { 
      \draw[orange,very thick] (0,\y)--(0.75,\y);
    }
  \end{scope}
  \end{scope}
  \end{scope}
\end{tikzpicture}
  \caption{(Color online) Degeneracies of the ground states as a
    function of  the divisor $\Np$ labeling the phases 
     of $\ZZ_8$ parafermions for both closed (left)
     and open (right) boundary conditions. In the closed chain the
     degeneracy is entirely due to symmetry breaking. The additional
     ground states (pale) in the open chain are accounted for by the
     appearance of symmetry protected edge modes.}
  \label{fig:para8}
\end{figure}

Before moving on, we discuss the implications for 
a system with periodic boundary conditions. Namely, in
  that case the ground state degeneracy due to topologically protected
  edge modes is expected to disappear while the degeneracy due to
  symmetry breaking should persist.
  The degeneracy due to symmetry breaking 
  is $\frg$, and therefore one expects a closed parafermionic
  chain to have a $\frg$-fold degenerate ground state. Whenever
  $\frg<\ml$ our arguments imply the existence of topologically
  protected edge modes.
We illustrate this phenomenology in the case of $\ZZ_8$ parafermions
in Figure \ref{fig:para8}, see also Table~\ref{tab:Overview} for a
more abstract summary of our findings. Our expectations will be verified in
Section \ref{sc:Realization} using the explicit realization of the
phases.

\subsection{\label{sc:Realization}Realization of the phases}

  In this section we will provide an explicit realization of all
  parafermionic phases discussed in the previous section, thereby
  putting our considerations on a concrete basis. It turns out to be
  convenient to discuss open and closed boundary conditions
  separately.
  We interpret a reduction in the ground state degeneracy when closing
  the chain as one of the signals for the presence of topological
  order.

\subsubsection{\label{sc:OpenBC}Open boundaries.}

The Hamiltonian for the open chain is
\begin{equation}
  \label{eq:Hop}
  \Hop = -\sum_{i=1}^{L-1}\hi \, ,
\end{equation}
  where $\hi$ is given in \eqref{eq:hiip1_para} for any fixed choice
  of phase $\Np$. $\Hop$ has $\ml$
  degenerate ground states $|\phi_\alpha\rangle\, ,\alpha\in \ZZ_\ml$,
  see Eq.~\eqref{eq:psia}. Part of this degeneracy is due to a
  symmetry breaking $\ZZ_N\to\ZZ_{N/\frg}$ and can be
  understood in terms of zero modes (operators commuting with the
  Hamiltonian $\Hop$) $\gamma_{j}^{p N/\frg}$ which are {\em
    delocalized} along the whole chain. Indeed, from
  Eq.~\eqref{eq:comm_gamma_nm} we see that $\bigl[\gamma_{j}^{p
    N/\frg},\Hop \bigr]=0$ for every $j$, and acting with powers of
  $\gamma_{j}^{N/\frg}$ on a reference ground state with definite
  parity, say $|\phi_0\rangle$, will generate a $\frg$-dimensional
  subspace in the space of ground states. Local perturbations can be
  used to select a specific direction and lift this degeneracy, see
  Eq.~\eqref{eq:SSB}.

When $\ml/\frg>1$ the open system exhibits additional ground states. All
of them can be obtained using the action of zero modes
$\gamma_1^{\eta\Np}$ and $\gamma_{2L}^{\eta\Np}$ (with
$\eta=1,\ldots,m/\frg-1$), which are {\em localized} at the
boundary. Note that the resulting edge states are distinguished by
their degree.
The degeneracy associated with the edge zero modes
is topological and robust against local perturbations.
  This is a consequence of the following observation to be established
  below: The presence of edge modes implies that the
  residual symmetry transformations in $\ZZ_{N/g}$ can be
  effectively factorized as the product of two operators localized
  respectively at the left and right boundary. The fractionalization
  of these symmetries protects the topological part of the ground
  state degeneracy and allows to define a topological
  invariant. In this sense the parafermionic phases fit into the
  general framework developed for classifying symmetry protected
  topological phases of spin chains with broken symmetries
  \cite{Schuch:1010.3732v3,Chen:PhysRevB.84.235128}.

  Before we comment on the general situation let us first of all stick
  to the setting of the fine tuned Hamiltonians \eqref{eq:hiip1_para}.
Consider the action of $P^g$, the generator of the unbroken
symmetries, on the ground state manifold:
$P^g |\psi_\alpha\rangle = |\psi_{\alpha+g} \rangle$.  (Recall that
also $P^m |\psi_\alpha\rangle = |\psi_{\alpha} \rangle$, since
$\alpha$ is defined modulo $\ml$.)  We postulate that this action can
be replaced by that of the operator $\hat{P}_g$ written in terms of
edge zero modes as
\begin{equation}
  \label{eq:Pg_fact}
  \hat{P}_g = 
  \omega^{\epsilon \Np(N+\epsilon \Np)/2} 
  \gamma_1^{\epsilon \Np} \gamma_{2L}^{-\epsilon \Np}\, ,
\end{equation}
with a well-defined, yet hitherto undetermined, number
$\epsilon\in\ZZ_{\ml/g}$ which only depends on the choice of divisor
$\Np$. The phase factor is chosen such that
$\hat{P}_g = 
  \sigma_1^{\epsilon \Np} 
  (\sigma_{L}^\dagger )^{\epsilon \Np} P^{\epsilon \Np}$,
from which we compute $\hat{P}_g |\psi_\alpha\rangle = 
  |\psi_{\alpha+\epsilon \Np}\rangle$. A comparison with the original
  expression $P^g|\psi_\alpha\rangle=|\psi_{\alpha+g}\rangle$
  shows that we can then determine the value of $\epsilon$ by imposing
\begin{equation}
    \label{eq:NpNpeps}
  \Np\epsilon = g \pmod{\ml}\, .
\end{equation}
Taking into account that $g=\gcd(m,n)$, this equation has a unique
solution for $\epsilon\in \ZZ_{\ml/g}$.
Indeed if $\epsilon_1=(g+p_1\ml )/\Np$ and $\epsilon_2=(g+ p_2\ml
)/\Np$ are two solutions, then $\epsilon_1-\epsilon_2 = (p_1-p_2)
\ml/\Np$ must be an integer, implying $(p_1-p_2)$ is a multiple of
$\Np/g$ and so $\epsilon_1-\epsilon_2 =0\pmod{\ml/g}$.  It is also
simple to show that the broken symmetries $\{P,\dots,P^{g-1}\}$ do not
admit a representation on ground states in the factorized form of
Eq.~\eqref{eq:Pg_fact}.
For $g=1$ the $\ZZ_N$ symmetry is not broken and the phase is purely
topological.

We have already seen that some of the parafermionic phases differ in
the symmetry group that is preserved by their ground states.
We next wish to establish that $x=\epsilon\Np\in\ZZ_{N/\frg}$ is
a topological invariant which can distinguish phases $\Np_1$ and
$\Np_2$ corresponding to the {\em same} symmetry breaking pattern
$\ZZ_N\to\ZZ_{N/\frg}$.
We note that the phases $\Np_1$ and $\Np_2$ come with a
different number of zero modes, $N/\Np_1$ and $N/\Np_2$,
respectively. One may hence argue that both phases are already
distinguishable. However, as we shall show in the following section,
the previous observation is an artifact of the open chain. In the
closed chain, both phases will exhibit $\frg$ ground states and hence
an independent invariant is required to discriminate
between the two. We have verified on the computer up to $N=5 \cdot
10^5$ that
two different phases $\Np_1\neq\Np_2$ (with the same $g$) never give
rise to the same invariant $x$.
At the moment we still lack an analytical proof for this
assertion.

For a fixed phase $\Np$, our definition yields a topological invariant
with values in $\ZZ_{N/\frg}$. However, comparing these
invariants for phases $\Np_1$ and $\Np_2$ with different values of
$\frg$ makes no sense, at least not a priori, since the preserved
symmetries protecting the non-trivial topology and therefore also the
groups of topological invariants are different. One could nevertheless
attempt to make them comparable by working with hierarchies of
topological phases, see \cite{Duivenvoorden:2012arXiv1206.2462D}.
In order to develop a better intuition for the topological
invariant, we collected the values of $x=\epsilon n$ for several types
of parafermions and phases in Table~\ref{tab:topinv}.

  Let us finally discuss whether our arguments rely on the particular
  choice of fine tuned Hamiltonians \eqref{eq:hiip1_para} or whether
  they remain true if one permits small but finite deformations. For
  the special case of Majorana chains ($N=2$) it has been established
  by Kitaev that the zero modes persist
  \cite{Kitaev:2001PhyU...44..131K}, at least in the limit of an
  infinite chain where a potential level splitting is exponentially
  suppressed. The same statement has been derived by Fendley for the
  purely topological phase of $\ZZ_N$ parafermions with $\Np=1$
  \cite{Fendley:2012arXiv1209.0472F}. While the extension of Fendley's
  argument to our more general setting is beyond the scope of our
  paper, it seems plausible that localized edge modes remain stable in
  a gapped system. If this is the case, the factorization of the
  generator $P^\frg$ of the $\ZZ_{N/\frg}$ symmetry on the low
  energy spectrum then again leads to a representation $\hat{P}_g=P_l
  P_r$ with $P_l$ and $P_r$ localized at the left and right boundary,
  respectively, but now over a distance of the correlation length (see
  \cite{Turner:2011PhRvB..83g5102T,Fidkowski:PhysRevB.81.134509} for
  the case of Majorana fermions). Just as before, this
  fractionalization is characterized by the $\ZZ_{N/g}$-degree $x$ of
  $P_l$. By its mere definition as an integer the value of $x$ cannot
  be changed smoothly and thus constitutes a $\ZZ_{N/\frg}$-valued
  topological invariant that can be used to characterize the (symmetry
  protected) topological phase. In the special case $N=2$ it reduces
  to the familiar expression for interacting Majorana fermions
  \cite{Turner:2011PhRvB..83g5102T,Fidkowski:PhysRevB.81.134509}.

\begin{table}
\centering
\begin{tabular}{cccccc} 
\toprule
& & & \multicolumn{2}{c}{Ground state degeneracy} \\ \cmidrule(r){4-5}
Type: $N$ & Phase: $\Np$ & Symmetry &
SSB: $\frg$ & Edge: $N/\Np\frg$ & Invariant: $x$ \\ \midrule
  \multirow{2}{*}{Prime $p$} &
  $\mathbf{1}$ & $\ZZ_p$ & $\mathbf{1}$ & $\mathbf{p}$ & $\mathbf{1}$ \\
  & $\mathbf{p}$ & $\ZZ_p$ & $\mathbf{1}$ & $\mathbf{1}$ & 
$\mathbf{0}$\\
\midrule
  \multirow{4}{*}{$8$} & $\mathbf{1}$ & $\ZZ_8$ & $\mathbf{1}$ & $\mathbf{8}$ & 
$\mathbf{1}$ \\    
  & $2$ & $\ZZ_4$ & $2$ & $2$ & $2$ \\    
  & $4$ & $\ZZ_4$ & $2$ & $1$ & $0$ \\    
  & $\mathbf{8}$ & $\ZZ_8$ & $\mathbf{1}$ & $\mathbf{1}$ & 
$\mathbf{0}$ \\    
\midrule
\multicolumn{1}{c}{\multirow{6}{*}{$12$}} &
\multicolumn{1}{c}{$\mathbf{1}$} & $\ZZ_{12}$ & $\mathbf{1}$ & $\mathbf{12}$ & 
$\mathbf{1}$ \\    
\multicolumn{1}{c}{} &
\multicolumn{1}{c}{$2$} & $\ZZ_6$ & $2$ & $3$ & $2$ \\    
\multicolumn{1}{c}{} &
\multicolumn{1}{c}{$\mathbf{3}$} & $\ZZ_{12}$ & $\mathbf{1}$ & $\mathbf{4}$ & 
$\mathbf{9}$ \\    
\multicolumn{1}{c}{} &
\multicolumn{1}{c}{$\mathbf{4}$} & $\ZZ_{12}$ & $\mathbf{1}$ & $\mathbf{3}$ & 
$\mathbf{4}$ \\    
\multicolumn{1}{c}{} &
\multicolumn{1}{c}{$6$} & $\ZZ_6$ & $2$ & $1$ & $0$ \\    
\multicolumn{1}{c}{} &
\multicolumn{1}{c}{$\mathbf{12}$} & $\ZZ_{12}$ & $\mathbf{1}$ & $\mathbf{1}$ & 
$\mathbf{0}$ \\    
\bottomrule
  \end{tabular}
  \caption{List of some $\ZZ_N$ parafermionic phases together with
    their properties. The number
    $\Np$ runs through the divisors of the parafermionic type
    $N$ and labels the possible phases,
    $\frg$ and $N/\Np\frg$ are respectively symmetry breaking and topological
    degeneracies of the ground state of a system with open boundaries,
    and $x$ is a topological invariant allowing to distinguish 
    different phases with the same broken symmetries.
    Rows in boldface correspond to
    purely topological phases ($\frg=1$).
  }
  \label{tab:topinv}
\end{table}

We now show that the existence of the phases realized by our
Hamiltonians \eqref{eq:hiip1_para} can be predicted on general grounds
by discussing in how many possible ways any potential residual
symmetry $\ZZ_{N/g}$ can be ``fractionalized''.
The appealing feature of this approach is that it does not make any
reference to a concrete Hamiltonian, thus giving additional support to the fact
that the classification developed so far is indeed complete. The
following discussion generalizes previous insights for Majorana chains
\cite{Turner:2011PhRvB..83g5102T,Fidkowski:PhysRevB.81.134509}, and
also a recent work on parafermions \cite{Motruk:2013arXiv1303.2194M}
which focuses on purely topological phases ($\frg=1$ in our
notation). Suppose again that the action on low energy states of the
symmetry transformation $P^g$ can be
represented effectively as $\hat{P}_g=P_l P_r$, with $P_l,P_r$
operators of definite $\ZZ_{N/\frg}$ degree acting respectively on the
left and right boundaries. Here $g$ is fixed and taken from the set
of possible values of $\gcd(n',N/n')$, where $n'$ is running through the
divisors of $N$. In the thermodynamic limit, the existence of
non-trivial factorizations implies topological degeneracies due to
edge modes, and hence allows to distinguish different topological
phases. We will next derive and solve consistency conditions for the
existence of such a factorization.

Denoting by $x$ the $\ZZ_{N/\frg}$ degree of $P_l$ (and by $-x$ that
of $P_r$), the consistency of the assigned grades with the
factorization requires that
\begin{align}
   \hat{P} P_l = \omega^{\frg x} P_l \hat{P}
   = P_l P_r P_l = \omega^{x^2} P_l \hat{P}\, .
\end{align}
In other words, the degree of $P_l$ must satisfy the equation
\begin{align}
  \label{eq:xx-1}
  x(x-\frg) = 0 \pmod{N}\, , \quad \text{with } x\in \ZZ_{N/\frg}\, .
\end{align}
The solutions to this equation label all non-equivalent factorizations
of $P^\frg$ and correspondingly all the possible topological phases
with a given pattern of symmetry breaking $\ZZ_N\to \ZZ_{N/\frg}$.

Note that Eq.~\eqref{eq:NpNpeps} can be rewritten as $\Np(\Np\epsilon
- g) = 0\pmod{N}$, so that any of its solutions gives rise to a solution
of Eq.~\eqref{eq:xx-1} with $x=\Np\epsilon$. In other words, the
factorizations \eqref{eq:Pg_fact}
discussed for our fine tuned Hamiltonians \eqref{eq:hiip1_para} are
indeed predicted by Eq.~\eqref{eq:xx-1}. In fact, we claim
that the solutions of Eq.~\eqref{eq:NpNpeps} realize all the possible
solutions to that equation. To support our assertion, we first
associate to each solution of Eq.~\eqref{eq:xx-1} a divisor $\Np_x$ of
$N$ with $\gcd(\Np_x,N/\Np_x)=g$. This can be achieved by setting
$\Np_x:=\gcd(x,N/\frg)$. The proof of this fact can be easily carried
out using simple properties of the greatest common divisor:
\begin{align}
  \gcd(\Np_x,N/\Np_x)
  &= \gcd\left(\gcd(x,N/\frg),\frac{N}{\gcd(x,N/\frg)} \right)
   = \frac{\gcd\bigl(\gcd(x,N/\frg)^2,N\bigr)}{\gcd(x,N/\frg)}\\
  &= \frac{\gcd\bigl(\gcd(x^2,(N/\frg)^2),N\bigr)}{\gcd(x,N/\frg)}
   = \frac{\gcd\bigl(x^2,\gcd\bigl((N/\frg)^2,N\bigr)\bigr)}{\gcd(x,N/\frg)}\\
  &\!\!\!\!\overset{\eqref{eq:xx-1}}{=} \frac{\gcd\bigl(x g\text{ (mod\,}N),N\bigr)}{\gcd(x,N/\frg)}=g\cdot
  \frac{\gcd(x,N/\frg)}{\gcd(x,N/\frg)}=g\, .
\end{align}
By the definition of $n_x$, the number $\epsilon_x:=x/n_x$
is an integer between $0$ and $N/(n_x g)-1$. The final step would
consist of establishing the relation
\begin{equation}
  \epsilon_x  n_x = g \pmod{N/n_x}\, ,
\end{equation}
which is exactly Eq.~\eqref{eq:NpNpeps}.  Unfortunately at present we
do not know how to prove this last equation, but its validity has been
checked numerically case by case up to $N=5 \cdot 10^5$.
Therefore the solutions of \eqref{eq:xx-1} can be expressed by those
of \eqref{eq:NpNpeps} in the form $x=\epsilon\Np$, and our
Hamiltonians \eqref{eq:hiip1_para} provide an explicit realization of
all the phases predicted by demanding a consistent factorization of
unbroken symmetries.

We conclude with some remarks on the number of distinct phases
which preserve the full symmetry $\ZZ_N$.  In this case solutions
of the factorization constraint are given by so-called unitary
divisors, namely divisors $\Np$ of $N$ such that
$g=\gcd(\Np,N/\Np)=1$.  In the prime factorization $N=\prod
p_i^{a_i}$, $\Np=\prod p_i^{c_i}$ is a unitary divisor if each $c_i$
is $0$ or $a_i$. For example the divisors of $N=4$ are $1,2,4$ but
only $1,4$ are unitary.
The number of parafermionic phases characterized by
purely topological order (i.e.\ without symmetry breaking)
is given by the number of unitary divisors. For
$N=1,2,3,\dots$ their numbers are
(Sloane's \href{http://oeis.org/A034444}{A034444})
\begin{equation}
  1, 2, 2, 2, 2, 4, 2, 2, 2, 4, 2, 4, 2, 4, 4, 2, 2, 4, 2, 4,
4,4,2,4,2,4,2,4,2,8,
 \dots
\end{equation}
These numbers can also be identified with $2^q$, where $q$ is the
number of different primes dividing $N$.

Actually, it can easily be shown that a similar counting is at work if
one considers a symmetry breaking
$\ZZ_N\to\ZZ_{N/\frg}$ with $\frg\neq1$. Indeed, our goal is
to enumerate all phases $\Np$ with a specific value of
$\frg=\gcd(\Np,N/\Np)$. Writing $N=\prod p_i^{a_i}$ and $\Np=\prod
p_i^{c_i}$ as in the previous paragraph, one easily finds $\frg=\prod
p_i^{\min(c_i,a_i-c_i)}$. All the values of $\Np$ which give rise to
the same value of $\frg$ thus arise from the original one by either
keeping the same value of $c_i$ or replacing it by $a_i-c_i$, factor
by factor.
As long as $c_i\neq a_i-c_i$ for all indices $i$ (as is the case for
$g=1$), this reasoning again yields~$2^q$ phases, the number of
unitary divisors of $N$. If however the equation $c_i=a_i-c_i$ is
satisfied for one or more of the indices~$i$, the exponent will be
reduced accordingly.

\subsubsection{\label{sc:ClosedBC}Closed boundaries.}

The parafermionic closed chain is defined by the local Hamiltonian
\begin{equation}
  \label{eq:Periodic}
  \Hcl^P = \Hop - \hL^P\, .
\end{equation}
In specifying $\hL^P$ we need to pay attention to the order of
indices since parafermions on different sites do not commute.
The substitution $(2i,2i+1)\to (2L,1)$ in the term
$\gamma_{2i}^{\beta \Np}\gamma^{-\beta \Np}_{2i+1}$ of
Eq.~\eqref{eq:hiip1_para} inverts the order of indices, and 
two different Hamiltonians are obtained depending on whether we
commute the parafermions before or after the replacement of indices.
Indeed performing the above substitution and then commuting
the parafermions gives
$\gamma_{2L}^{\beta \Np}\gamma^{-\beta \Np}_{1}=\omega^{(\beta \Np)^2}
\gamma^{-\beta \Np}_{1}\gamma_{2L}^{\beta \Np}$.
If instead we commute the parafermions in Eq.~\eqref{eq:hiip1_para}
before replacing the indices we will get
$\omega^{-(\beta \Np)^2}\gamma^{-\beta \Np}_{1} \gamma_{2L}^{\beta \Np}$.
The ambiguous extra phase produced by inverting the operations
signals an ambiguity in the definition of $\hL^P$.
The correct definition of the coupling 
instead keeps the order of labels and implements the replacement
$(2i,2i+1)\to (1,2L)$ in $\gamma_{2i}^{\beta \Np}\gamma^{-\beta
  \Np}_{2i+1}$. Following this prescription we get
\begin{align}
  \label{eq:hL1_para}
  \begin{split}
  \hL^P&=\frac{1}{\ml}\sum_{\beta=0}^{\ml-1}
    \omega^{\beta \Np(N + \beta \Np)/2} 
  \gamma_{1}^{\beta \Np}\gamma^{-\beta \Np}_{2L}\\
&\qquad\qquad  \times \frac{1}{\Np^2}
  \sum_{j,k=0}^{\Np-1}
  \omega^{(\ml^2(j^2+k^2)-N\ml (j+k))/2}
  \gamma_{2L-1}^{-j\ml}
  \gamma_{2L}^{j\ml} 
  \gamma_{1}^{-k\ml}\gamma_{2}^{k\ml}\, .
  \end{split}
\end{align}
  When rewritten in terms of spins, this operator turns out to be
  non-local. More precisely, it contains a twisting by the inverse
  global parity $P^\dagger$:
\begin{align}
  \hL^P&=
  \frac{1}{\ml \Np^2}
  \sum_{\beta=0}^{\ml-1}(P^\dagger
    \sigma_L\sigma^\dagger_{1})^{\beta \Np} 
  \sum_{j,k=0}^{\Np-1}\tau_L^{j\ml}\tau_{1}^{k\ml} \, .
\end{align}
  While completely equivalent when regarded with open boundary
  conditions, the spin model and the parafermionic chain thus exhibit
  essential differences when considered on a closed ring.

  In order to understand the physical implications of our previous
  comment, we now study the spectrum of the periodic chain, with
  particular attention to the ground state and its degeneracy.  Recall
  that $\Hop$ has $\ml$ ground states which are given by
  $|\phi_\alpha\rangle$ as defined in Eq.~\eqref{eq:psia}. Since
  $[\Hop , \hL^P]=0$, the operators $\Hop$ and $\hL^P$ admit a common
  basis of eigenvectors, and our problem simply amounts to
  diagonalizing the projector $\hL^P$ on the space spanned by the
  vectors $|\phi_\alpha\rangle$.

  A brief calculation shows that the operator $\hL^P$ acts on the
  ground states of the open chain as
\begin{align}
  \label{eq:hpL1_psi}
  \hL^P|\phi_\alpha\rangle &=
  \frac{1}{\ml}\sum_{\beta=0}^{\ml-1}\omega^{\beta \Np^2 \alpha}
  |\phi_\alpha\rangle=
  \delta_{\alpha \Np=0}^{(\text{mod}\,{\ml})}|\phi_\alpha\rangle\, .
\end{align}
  Put differently, the state $|\phi_\alpha\rangle$ remains in the
  ground state manifold of the closed parafermionic chain if and only
  if
\begin{equation}
  \label{eq:alpha_per}
  \alpha \Np=0 \pmod{\ml}\, .
\end{equation}
  In order to enumerate the solutions to this equation, let us first
  introduce the symbols $\tilde{\Np}=\Np/\frg$ and
  $\tilde{\ml}=\ml/\frg$ where we used the previous abbreviation
  $\frg=\gcd(\Np,\ml)$. With this notation, we can write the solutions of
  Eq.~\eqref{eq:alpha_per} as
\begin{equation}
  \label{eq:AlphaVal}
  \alpha = p  \frac{\tilde{\ml}}{\tilde{\Np}}
  \quad\text{ with }\quad
  p\in\tilde{\Np}\ZZ_\frg\, .
\end{equation}
  Hence, the possible values of $p$ are
  $p=0,\tilde{\Np},\dots,\tilde{\Np}(\frg-1)$, so that there are $\frg$
  distinct solutions.

  It may easily be shown that the previous $\frg$ states exhaust the
  ground state manifold of the periodic Hamiltonian. Note that we can
  associate the degeneracy of the closed chain to parafermionic zero
  modes as in the case of the open chain. Indeed
  Eq.~\eqref{eq:comm_gamma_nm} implies that $\Bigl[\gamma_{i}^{p
    N/\frg},\Hcl^P \Bigr]=0$, and acting with powers of
  $\gamma_{i}^{N/\frg}$ on a reference ground state with definite
  parity will produce $\frg$ mutually orthogonal ground states which
  can be distinguished by their parity eigenvalues.

  Our analysis implies that the closed parafermionic chains have a
  unique ground state for $\Np$ being a unitary divisor, while there
  exists a $\frg$-fold degenerate ground state when $\Np$ fails to be a
  unitary divisor. In the former case, there is no spontaneous
  symmetry breaking and the ground state degeneracy in the open chain
  may be completely attributed to the non-trivial topology of the
  system. On the other hand, in the latter case, the ground state
  degeneracy is only partially lifted when making the transition from
  open to closed boundary conditions. In that case, we observe a
  combination of non-trivial topology and spontaneous symmetry
  breaking if $\ml/\frg>1$. For $\frg=\ml$ there is no non-trivial
  topology and only symmetry breaking. These results are in complete
  agreement with our expectations from Section \ref{sec:par_phases}.

  Phases of the closed parafermionic chain with the same symmetry
  breaking pattern and the same number of ground states can be
  characterized and distinguished by the topological invariant defined
  in Section~\ref{sc:OpenBC}. Even though the definition of the
  topological invariant requires an open chain, it should be clear
  that the latter can be realized virtually using the tool of
  entanglement spectroscopy \cite{Turner:2011PhRvB..83g5102T}, i.e.\
  by dividing the closed chain into two segments and tracing out the
  degrees of freedom associated with one of them.

\begin{table}
\begin{center}
\begin{tabular}{cc|cc}
  \multicolumn{2}{c}{Phase: $\Np$ (a divisor of $N$)} & $\ZZ_N$ spin model & $\ZZ_N$ parafermions
  \\\hline\hline&&\\[-1em]
  \multicolumn{2}{c|}{Symmetry breaking} & $\ZZ_N\to\ZZ_\Np$ & $\ZZ_N\to\ZZ_{N/\frg}$ \\
  \multicolumn{2}{c|}{Topological invariant} & $\emptyset$ & $x=\epsilon n\in\ZZ_{N/\frg}$\\
  \multirow{2}{*}{Ground state degeneracy
      $\begin{cases}\\[2mm]\end{cases}$\!\!\!\!\!\!\!\!\!\!\!\!\!\!} & open chain & $m=N/\Np$ & $g\cdot(N/\Np\frg)$ \\
  & closed chain & $m=N/\Np$& $\frg$
\end{tabular}
  \caption{\label{tab:Overview}Classification of gapped phases in
    $\ZZ_N$ spin models and in the dual parafermionic
    theories. In both cases, the phases are labeled by a divisor $\Np$
    of $N$ but the physical interpretation is rather different since
    the two theories are non-locally related. In particular, the
    ground state degeneracy in the open parafermionic chain factorizes
    into contributions from spontaneous symmetry breaking ($\frg$) and
    from topologically protected edge modes ($N/\Np\frg$). We also
    note that the closed chains are not equivalent since periodic
    boundary conditions lead to different Hamiltonians for the spin
    chain and the parafermions. The number $\frg$ is determined by
    $\frg=\gcd(\Np,N/\Np)$ while the constant $\epsilon$ is the unique
    solution to the equation $n\epsilon=\frg\ (\text{mod}\,N/\Np)$, see
    Eq.~\eqref{eq:NpNpeps}.}
\end{center}
\end{table}

\section{Conclusions and outlook}

  In the current paper we have classified the massive phases of
  $\ZZ_N$ parafermionic chains. This was achieved by constructing
  all gapped symmetry broken phases in the dual quantum spin chains. The
  number of distinct phases and their specific characteristics depend
  crucially on the number--theoretic properties of $N$. First of all,
  it is easy to construct one SSB phase of the spin model for each
  divisor $\Np$ of $N$. These phases are characterized by the
  (co-)existence of local and non-local order parameters. Since
  $\ZZ_N$ does not lead to symmetry protected topological phases
  \cite{Chen:PhysRevB.83.035107,Chen:PhysRevB.84.235128}, the
  enumeration of divisors exhausts all possibilities for gapped phases.

  In a second step we have then interpreted the resulting ground
  states and Hamiltonians from a parafermionic perspective. Of course,
  this leads to the same number of phases. However, in contrast to the
  spin model the parafermionic phases can now exhibit features of
  both, symmetry breaking and non-trivial topology. Whether the
  resulting parafermionic phase should be interpreted as topological
  first of all depends on whether $\Np$ and $N/\Np$ still have common
  divisors.  A parafermionic phase is purely topological if $\Np$ is a
  unitary divisor of $N$. By definition, this is the case if $\Np$ and
  the quotient $N/\Np$ are co-prime or, in our notation, if
  $\frg=\gcd(\Np,N/\Np)=1$. The number of unitary divisors is known to
  be $2^q$ where $q$ is the number of distinct prime factors in $N$.
  In contrast, a parafermionic phase described by $\Np<N$ not being a
  unitary divisor (i.e.\ $\frg>1$) definitively exhibits spontaneous
  symmetry breaking. If, in addition, $\frg\Np<N$ then the phase is
  also protected by topology.  Topology protected phases in open
  systems exhibit gapless parafermionic edge modes. These edge modes
  are gapped out when the chain is closed, while a potential
  degeneracy due to symmetry breaking persists.  We also proposed a
  topological invariant characterizing the topological properties of
  open chains. Table~\ref{tab:Overview} provides a compact, yet
  exhaustive, summary of our findings.

  While our current results are only concerned with the intrinsic
  $\ZZ_N$ symmetry of parafermionic chains, it would be
  interesting to extend our analysis to systems which are required to
  respect additional symmetries such as inversion or time-reversal
  symmetry. Alternatively, one could also add internal degrees of
  freedom which transform under a discrete or continuous symmetry
  group. In all these cases one is lead to so-called symmetry
  protected topological phases which, for bosonic systems, have been
  fully classified in Refs.\
  \cite{Chen:PhysRevB.83.035107,Schuch:1010.3732v3,Chen:PhysRevB.84.235128}.
  It should be noted that models with continuous symmetry lead to a
  few peculiarities that have been addressed in
  Ref.~\cite{Duivenvoorden:2012arXiv1206.2462D}.
  
  For fermionic chains it is known that the inclusion of additional
  symmetries has profound consequences. Most importantly, the
  restriction to time-reversal invariant systems enhances the group of
  topological invariants to $\ZZ$ or $\ZZ_8$, respectively,
  depending on whether interactions are allowed
  \cite{Fidkowski:PhysRevB.81.134509,Fidkowski:2011PhRvB..83g5103F,Turner:2011PhRvB..83g5102T}
  or not \cite{Ryu:1367-2630-12-6-065010,Kitaev:2009mg}. A general
  mathematical formalism for treating fermionic symmetry protected
  topological systems has been developed in
  Ref.~\cite{Gu:2012arXiv1201.2648G} and it appears feasible to
  extend it to the $\ZZ_N$-graded generalization of the Clifford
  algebra \cite{Morris1967} needed to describe parafermions.

  Finally, the classification scheme used in this paper is, of course,
  not limited to $\ZZ_N$ spin models. It should rather be
  applicable to any model of statistical physics which exhibits a
  Kramers-Wannier type of duality. While abelian dualities are
  well-explored (see, e.g., \cite{Savit:1980RvMP...52..453S}) and
  close to our exposition, recent progress on non-abelian dualities
  \cite{Cobanera:2012arXiv1206.1367C} might offer a vast playground
  for the exploration of new topological phases.
  
  Needless to say, knowledge about the classification of topological
  phases, symmetry
  protected or not, and a representative Hamiltonian for each phase
  should merely be the starting point for a more thorough
  investigation. Indeed, the ultimate goal should be to gain control
  over the complete phases diagram of physically realistic
  Hamiltonians, including the location and the 
  nature of phase transitions. For the Majorana chain with
  interactions such an analysis was provided in
  \cite{Hassler:2012NJPh...14l5018H}, based on the known phase diagram
  of the dual anisotropic next-nearest-neighbor Ising (ANNNI)
  model. It would be interesting to extend this analysis to the case
  of $\ZZ_N$ spin models. Also, with regard to the potential
  physical realization and detection of parafermions one will need to
  understand the effects of adding disorder to the couplings.

\begin{center}
  \em $\ast\ast\ast$\quad Note\quad $\ast\ast\ast$
\end{center}
  \vspace{-.8em}
  While in the process of completing this work, the paper
  \cite{Motruk:2013arXiv1303.2194M} appeared which has
  considerable overlap with our own results. We gratefully acknowledge
  useful discussions with J.\ Motruk and A.\ Turner about aspects of
  entanglement and symmetry fractionalization in parafermionic chains
  prior to the publication of their paper.


\ack{
  We would like to thank A.\ Altland for interesting discussions
  leading to the launch of this project and K.~Duivenvoorden for
  collaboration in the initial stages and extremely
  useful discussions.
  Both authors are funded by the
  German Research Council (DFG) through M.\ Zirnbauer's Leibniz Prize,
  DFG grant no.\ ZI 513/2-1. Additional support is received from the DFG
  through the SFB$|$TR\,12 ``Symmetries and Universality in Mesoscopic
  Systems'' and the Center of Excellence ``Quantum Matter and
  Materials''.
}

\section*{References}


\providecommand{\href}[2]{#2}\begingroup\raggedright\endgroup

\end{document}